# Quantum Standards & Accounting Information Systems

Maksym Lazirko
Department of Accounting & Finance, Montclair State University
For all inquiries, please email lazirkom@montclair.edu
1 Normal Avenue , Montclair, NJ 07043, United States

**Abstract-**
Recent advancements in quantum technology threaten the cryptographic foundations of Accounting Information Systems (AIS), necessitating a transition to quantum-safe standards. This paper investigates why quantum standards fall within the purview of accounting by framing them as essential institutional governance mechanisms that ensure the integrity, auditability, and legitimacy of data. Utilizing neo-institutional theory, the study analyzes how coercive, normative, and mimetic pressures drive the adoption of these standards across jurisdictions. Through a structured documentary analysis of major standard-setting bodies, the research identifies significant divergence between U.S. and EU/European approaches: U.S. standards emphasize market-driven innovation and pragmatic legitimacy, while EU and Pan-European standards prioritize regulatory harmonization and societal privacy objectives. The findings suggest that while these standards are currently voluntary, their inconsistent implementation creates risks of decoupling and fragmented assurance practices, challenging the global comparability of AIS security controls.

---

## I. Introduction

Current accounting information systems (AIS) cybersecurity frameworks and assurance practices were largely designed around classical cryptographic threats, raising concerns about whether they can continue to provide comparable assurance as quantum-enabled attacks emerge. In most jurisdictions, the standards designed to protect AIS against quantum threats are voluntary, and the organizations they are meant to govern are under no binding obligation to adopt them. Recent advances in quantum computing threaten to undermine the cryptographic primitives that safeguard contemporary AIS, exposing data, technology-enabled controls, and assurance processes to novel classes of cyber risk (Kahyao, 2023). In response, a rapidly expanding but fragmented set of quantum-related standards is emerging from standards-setting bodies, yet accounting research has not systematically examined how this evolving quantum standards infrastructure will shape AIS design, governance, and auditability. This gap is problematic for accounting because differences in the scope, authority, and enforcement of quantum standards can generate cross-jurisdictional inconsistencies in how firms harden AIS against quantum-enabled attacks, document compliance, and signal reliability to capital market participants. Building on theories of disclosure regulation that show how accounting standards emerge endogenously from institutional bargaining and may diverge from diversified investors' preferences, quantum standards are likely to interact with existing frameworks in ways that may either complement or conflict with established AIS control architectures (Bertomeu & Cheynel, 2013). Accordingly, this study asks: *How do emerging quantum standards differ in their scope, governance, and enforcement mechanisms, and what are the implications of these differences for the design, control, and assurance of accounting information systems?*

Quantum computing is transitioning from a largely theoretical construct to an applied technology capable of breaking widely used public-key schemes such as Rivest-Shamir-Adleman (RSA), Diffie-Hellman, and elliptic-curve (EC) cryptography, thereby directly threatening the confidentiality and integrity of data processed and stored in AIS (Hossain Faruk et al., 2022). In parallel, AIS cybersecurity literature has developed along four main streams: risks and threats, controls, cybersecurity-related assurance, and breaches but has only begun to incorporate quantum-specific attack and defense vectors into its frameworks (Cram et al., 2023). As shown in the upcoming sections, standard setters and consortia in the United States are now issuing post-quantum cryptography (PQC), quantum key distribution (QKD), and broader quantum technology standards, almost all on a voluntary basis and oriented toward heterogeneous stakeholder groups. These quantum standards coexist with, and sometimes cut across, financial reporting, assurance, and information technology (IT) control regulations. Firms must decide when and how to retrofit AIS architectures, adopt quantum-safe algorithms, and evidence compliance to auditors and regulators.

Standards issued by standard setting organizations operate as institutional pressures: coercive, normative, and mimetic that influence how firms design, secure, and audit their AIS in anticipation of quantum-enabled threats (DiMaggio & Powell, 1983). Organizations adopt standards both to align with coercive, normative, and mimetic pressures and to signal technological sophistication, cybersecurity posture, and compliance quality to auditors, regulators, and investors. When audits or other evaluations uncover decoupling between formal claims of adherence and underlying AIS practices, the resulting loss of legitimacy can erode stakeholder trust. Framing quantum standards as institutionalized governance mechanisms and signaling devices enables the study to theorize how variations in the quantum standards infrastructure give rise to heterogeneous AIS architectures, control configurations, and assurance practices across jurisdictions.

Methodologically, the study conducts a structured documentary analysis of quantum computing related standards and guidance issued by major U.S. and European standard setters and consortia, complemented by a targeted review of AIS cybersecurity research. Candidate organizations are first identified via systematic web searches, after which their public standards, technical reports, and program descriptions are collected and coded along dimensions implied by the institutional and signaling lens scope, and explicit references to accounting relevant use cases. This coding produces a comparative matrix of quantum standards organizations (exemplified in Table 2) that is then interpreted against the four AIS cybersecurity research streams to infer likely consequences for AIS architecture, control design, and audit evidence generation across U.S. and European contexts.

The analysis reveals that the contemporary quantum standards are populated by a dense network of predominantly U.S. and European organizations that have only recently intensified their quantum-related activities, broadly converging in the 2010s and especially the 2020s as quantum technologies moved from experimental to applied contexts. Across these bodies, commonalities such as shared emphasis on post-quantum cryptography, quantum key distribution, and secure quantum communications and pronounced differences in regulatory orientation, application focus, governance, and funding are documented, with European efforts more tightly coupled to harmonization, privacy, and societal objectives and U.S. efforts more explicitly shaped by market-driven innovation and industry engagement. The findings further indicate that quantum standards are usually voluntary, unevenly adopted, and evolving more slowly than the underlying technologies, which creates fragmentation in guidance, potential gaps in enforceability, and nontrivial integration challenges with existing regulatory regimes. For AIS, this implies that early adopters operate in a heterogeneous and still-evolving standards environment in which transitioning to quantum-safe cryptography entails material implementation costs, expertise and training demands, performance and energy-efficiency trade-offs, and difficult judgments about long-term algorithm viability and compliance signaling to auditors, regulators, and investors.

This study provides one of the earliest structured institutional mappings of the quantum standards ecosystem explicitly anchored in AIS concerns, documenting which U.S. and European bodies set which quantum standards, for which stakeholder groups, and with what authority and enforcement characteristics, and showing how these institutional differences shape the pressures faced by AIS designers, controllers, and auditors. Integrating a neo-institutional and signaling lens with the four established AIS cybersecurity research streams, the paper develops a conceptual framework that links specific features of quantum standards such as voluntariness, governance structure, scope, and certification mechanisms to concrete implications for AIS architecture, control design, and audit-evidence generation across U.S. and European settings. The resulting comparative matrix of standard-setting organizations and their quantum standards offers practitioners and researchers an analytically grounded basis for prioritizing quantum-safe investments, interpreting firms' claims of compliance with quantum standards, and anticipating where misalignment between emerging quantum standards and existing financial reporting and IT-control regimes is likely to produce fragmented assurance practices and decoupling risks in AIS environments. The following sections include §II a review of the literature on accounting standards, institutional theory, and the AIS cybersecurity research streams relevant to quantum computing; §III an analysis of the major quantum standard-setting organizations and their standards, including a comparative assessment of U.S. and Pan-European orientations, adoption pathways, cross-body differences and similarities, and current limitations; §IV a discussion of implications and challenges for AIS design, control, and assurance; and a conclusion that distills recommendations and directions for future research.

## II. Literature Review

### *2.1 Scope & Relevance to Accounting*

Accounting standards constitute a core component of the institutional architecture of financial reporting, structuring how organizations construct and communicate economic reality and thereby shaping comparability, transparency, and legitimacy in the eyes of investors and other stakeholders across jurisdictions.

In a highly institutionalized accounting field, the diffusion of a common set of standards such as the International Financial Reporting Standards (IFRS) reflects pressures that promote convergence toward a shared global "financial reporting language." This enhances financial statement comparability, facilitates cross-border investment, and mitigates information asymmetry between firms and capital providers, thereby improving the overall information environment at the level of the organizational field (Schiavi et al., 2024). Huang and Yan (2020) analytically demonstrate that greater financial statement comparability lowers firms' cost of capital only when accounting standards are of sufficiently high quality, underscoring the complementary roles of comparability and standard quality. Daske et al. (2008) provide early evidence from mandatory IFRS adoption in 26 countries, documenting increases in market liquidity and reductions in the cost of capital (and higher equity valuations) around IFRS adoption once anticipation effects are accounted for, with these capital-market benefits arising only in countries where firms have incentives to be transparent and where legal enforcement is strong (Huang & Yan, 2020; Daske et al., 2008).

Several persistent debates animate the accounting standards literature. The first concerns mandatory versus voluntary adoption. Houqe (2018) finds that enforcement is "a recurring theme throughout the literature" on IFRS adoption and argues for greater international collaboration between the IASB and regulatory bodies to maximize standard effectiveness (Houqe, 2018). Li and Yang (2016) find that mandatory IFRS adoption significantly increases firms' issuance of management earnings forecasts, implying that mandatory reporting standards can spill over to reshape firms' broader voluntary disclosure incentives (Li & Yang, 2016). The second debate involves principles-based versus rules-based accounting. For instance when analyzing Portugal's shift from a code-law, rules-based institutional logic to an IFRS-based, principles-oriented regime, findings show that the rule-based practices embedded in the old code-law logic limited preparedness and fostered resistance to change. Even as firms voluntarily adopted the new system to sustain social legitimacy, they did so often in ways that remained loosely coupled to entrenched rules-based routines (Soares Fontes et al., 2021). Lysak (2020) provides evidence from the other direction. When she examined Big Four comment letters to the FASB, Lysak finds that extensive lobbying and uncertainty language from auditors are associated with increases in rules-based attributes in final standards, suggesting that auditor litigation risk preferences push standard-setters toward more prescriptive rules (Lysak, 2020). This is connected to the third debate: the political economy of standard-setting. Monsen (2022) provides systematic evidence that Big Four lobbying positions reflect profit motives, favoring proposed standards that are expected to generate additional fees or are supported by their clients, and that these positions are significantly associated with FASB standard-setting outcomes, both on a stand-alone basis and relative to other constituents, including financial statement users (Monsen, 2022). Allen, Ramanna, and Roychowdhury (2018) examine Big N auditors' comment letter lobbying on U.S. GAAP over the first 33 years of the FASB (1973-2006) found that lobbying is associated with prevailing litigation and regulatory scrutiny, but they find no evidence that lobbying is driven by client preferences for flexibility in GAAP (Allen et al., 2018). Kalaitzake (2019) argues that the Big Four function as "key political allies of the financial sector" in regulatory battles, documenting how, in the EU Financial Transaction Tax debate, they disseminated oppositional claims, prepared negative impact assessments, and advised financial firms on lobbying tactics (Kalaitzake, 2019).

Collectively, these debates establish that accounting standards constitute contested institutional arrangements shaped by economic incentives, political power, and stakeholder demands. This framing renders their analysis simultaneously theoretical and practical.

### *2.2 Institutional Theory (Neo-institutionalism)*

Neo-institutional theory provides the analytical framework for interpreting the standard-setting contests, legitimacy-seeking behaviors, and isomorphic pressures documented in the preceding debates. Developed in organizational studies in the late 1970s and 1980s as a response to explanations of organizational behavior grounded primarily in rational efficiency, neo-institutionalism through the foundational contributions of Meyer and Rowan (1977) and DiMaggio and Powell (1983) advanced the view that organizations pursue legitimacy by conforming to prevailing institutional norms, rules, and expectations, rather than solely by optimizing technical performance (Meyer & Rowan, 1977; DiMaggio & Powell, 1983).

Within this framework, institutions are conceptualized as the "rules of the game" that structure social action, encompassing formal regulation, informal conventions, and cultural-cognitive schemas that render certain practices taken for granted. In accounting, neo-institutionalism has been used to explain how practices such as standard setting, adoption and reporting are shaped by isomorphic pressures that encourage organizational similarity as a means of coping with uncertainty and securing survival. Legitimacy is central; organizations signal appropriateness and reduce scrutiny by conforming to institutionalized expectations. In contrast "old" institutionalism emphasized habit and inertia while neo-institutionalism stresses active processes of institutionalization through which organizations both respond to and reproduce institutional pressures. This perspective has been especially influential in accounting research examining the diffusion of standards such as IFRS, emphasizing their role as organizational responses to institutional environments that privilege legitimacy, conformity, and field-level stability (Schiavi et al., 2024) .

The concepts of isomorphism, legitimacy and institutional logics are mutually reinforcing. Isomorphic pressures generate legitimacy through conformity, while institutional logics provide the interpretive frame within which such pressures are perceived and negotiated. In accounting research, they are combined to illuminate how firms reconcile technical and institutional demands, as further elaborated in the applications discussed below. In the sustainability accounting literature, coercive, mimetic and normative pressures are widely found to underpin the adoption of environmental accounting standards, particularly in emerging economies and in the case of local government these institutional pressures operate alongside strong contingency-type organizational drivers (Qian et al., 2011). Overall, isomorphic pressures promote enhanced accountability, but their influence is contingent on local institutional contexts; for example, mimetic emulation of global leaders can accelerate standard adoption, yet decoupling may occur where local logics are misaligned with imported practices (Siti‑Nabiha & Scapens, 2005; Kostova et al., 2008).

### *2.3 Supplementary Theories to Neo-Institutionalism*

Building on the neo-institutional perspective, several alternative theories offer more actor-centered explanations of CSR, voluntary disclosure, and environmental accounting practices. Whereas neo-institutionalism emphasizes field-level isomorphic pressures, these approaches foreground how firms strategically manage relationships with specific audiences and resolve competing interests.

Legitimacy theory explains Corporate Social Responsibility (CSR) and environmental disclosures as tools for maintaining congruence between corporate behavior and societal expectations, thereby securing continued approval and pre-empting legitimacy threats. In this view, managers exercise agency by deploying reports and external assurances as symbolic demonstrations of accountability (Greenwood et al., 2008). This emphasizes the strategic and communicative use of reporting within broader institutional constraints, complementing rather than replacing the more structural accounts of conformity and isomorphism developed in neo-institutional theory. Stakeholder theory extends the analysis by conceptualizing firms as embedded in networks of primary and secondary stakeholders whose power, legitimacy, and urgency shape disclosure priorities, thus mediating institutional pressures through relational dynamics rather than treating them as diffuse environmental forces (Freeman et al., 2010). Stakeholder-agency theory further integrates agency theory with stakeholder considerations, positing that CSR disclosures help reduce information asymmetries and align interests across a wider set of principals and agents, offering a micro-level complement to the macro-level focus of neo-institutionalism (Del Gesso & Lodhi, 2024).

Across empirical accounting studies, these theories consistently show that stakeholder pressures and legitimacy concerns are associated with more extensive and, in some cases, higher-quality disclosures, especially in voluntary regimes. Evidence from settings such as France, Germany, the UK, and emerging economies indicates that organizational visibility, ownership structures and stakeholder salience condition the extent to which firms respond to, or go beyond, institutional requirements in their reporting (Fifka, 2013; Gamerschlag et al., 2011). In this way, legitimacy, stakeholder and stakeholder-agency theories do not replace neo-institutionalism; rather, they refine it by specifying how concrete stakeholder relationships and evaluations translate broad institutional norms into disclosure strategies and practices. Taken together, these perspectives suggest that standard adoption is rarely a purely technical choice, it reflects institutional pressures, stakeholder evaluations, and power relations. The same dynamics are visible in quantum standards, where voluntary frameworks, professional norms, and industry coalitions structure how organizations interpret and implement new requirements. In what follows, we extend this theoretical lens to the emerging ecosystem of quantum standards that condition the design, security, and auditability of AIS. Table 1 maps the four AIS cybersecurity research streams identified by Cram et al. (2023) onto the quantum cybersecurity, illustrating how each stream's core focus, key AIS topics, and quantum computing implications converge to define the research frontier addressed in this paper. While there are other parts of quantum technology that may interact with accounting , cybersecurity is the most obvious to dissect as a case example due to extensive prior research in AIS.

| Research Stream | Core Focus | Key AIS Topics | Quantum Implications |
|---|---|---|---|
| Cybersecurity Risks & Threats | Examines the nature, sources, and organizational impact of threats to AIS confidentiality, integrity, and availability, including malware, insider threats, phishing, and ransomware. | Threat taxonomy, vulnerability assessment, attack vectors, financial data exposure, ransomware, phishing, and insider threats | Quantum computers can break current public-key encryption (RSA, ECC), exposing AIS to harvest-now-decrypt-later attacks and requiring post-quantum threat modeling |
| Cybersecurity Controls | Focuses on technical and organizational mechanisms protecting AIS, including access controls, encryption, multi-factor authentication (MFA), and security policies | Internal controls, IT governance, encryption standards, multi-factor authentication (MFA), and security policy design | Legacy encryption controls (AES-128, RSA-2048) become inadequate against quantum adversaries; NIST post-quantum standards (CRYSTALS-Kyber, CRYSTALS-Dilithium) require wholesale control redesign |
| Cybersecurity-Related Assurance | Examines how auditors and governance bodies assess the adequacy of cybersecurity controls and communicate cyber risk to stakeholders | IT audit, SOC reports, cybersecurity disclosure standards, regulatory compliance, and | Quantum-safe standard adoption creates new assurance gaps; auditors lack |

| Research Stream | Core Focus | Key AIS Topics | Quantum Implications |
|---|---|---|---|
| | | audit committee oversight | established frameworks for evaluating post-quantum readiness in AIS environments |
| Cybersecurity Breaches | Analyzes the incidence, disclosure, and financial and reputational consequences of data breaches in AIS and financial reporting contexts | Breach notification, incident response, regulatory penalties, financial loss quantification, and stakeholder impact disclosure | Quantum-enabled breaches may remain undetected for extended periods; post-breach disclosure standards do not yet account for quantum facilitated intrusion vectors or harvest now decrypt later exposure |

**Table 1: AIS Cybersecurity Research Streams & Quantum Computing Implications (Cram et al., 2023)**

## III. Quantum Computing Standards

### *3.1 Overview of Quantum Standard Setters*

The complexity and diversity of quantum applications render a one-size-fits-all regulatory approach impractical. Instead, standards shaped by normative expectations within professional and technical communities and mimetic tendencies among industry participants offer a flexible framework that encourages participation, innovation, and adaptive alignment across diverse organizational contexts (Johnson, 2019). While mandatory regulations risk stifling innovation by imposing uniform coercive pressures that may not suit emerging quantum technologies, voluntary standards enable stakeholders to respond to a blend of institutional influences, adopting practices that fit their specific capabilities, resource dependencies, and strategic goals (Lim & Prakash, 2014). In this way, they function as a repository of recognized best practices, guiding organizations toward commonly accepted technical norms and fostering patterns of convergence through well-tested approaches (Rosen et al., 2002).

Organizations that embrace quantum standards[1] often gain a competitive edge by signaling legitimacy and alignment with broader institutional expectations. Companies demonstrating adherence to recognized frameworks project commitment to quality, security, and compliance, thereby enhancing reputation, building trust with clients and investors, and attracting partners through visible isomorphism with industry leaders (Alzeban, 2019). Given that quantum technologies frequently incorporate intricate cryptographic methods and sensitive data, following established standards helps mitigate risks of security breaches or system vulnerabilities by embedding practices that respond to both coercive regulatory signals and normative professional benchmarks. Adherence to social, environmental, and governance (ESG) reporting standards can signal a commitment to accurate and credible disclosures, and, when such compliance is validated through audit processes, it can help maintain or enhance organisational legitimacy, which may, in turn, support stakeholder confidence (Deegan, 2002). Furthermore, shared standards promote interoperability and collaboration by encouraging mimetic alignment, making systems more compatible, easing integration challenges, and accelerating the diffusion of quantum technologies across networks (Pitwon & Lee, 2021).

Voluntary standards can also serve as a bridge where quantum technologies intersect with formal regulatory requirements. By adopting them, organizations demonstrate proactive responses to coercive institutional pressures, minimizing legal and compliance risks while aligning practices with evolving governance structures (Chowdhury, 2013). Many such standards incorporate ethical considerations and responsible innovation norms. Standard uptake signals dedication to societal legitimacy and stakeholder expectations, resonating with consumers, investors, and other actors who value ethical alignment (Valentine & Fleischman, 2008). Adherence to these frameworks future-proofs organizations by equipping them to maintain legitimacy amid technological change. In accounting contexts, reports referencing compliance with quantum standards should signal transparent and reliable financial practices; conversely, discrepancies uncovered in audits between stated adherence and actual behavior can erode legitimacy, undermine investor trust, and damage organizational credibility (Hoang & Phang,

[1] Just like accountants follow GAP or IFRS, there are local and international standards! For example in the cybersecurity world, NIST is like GAP and ISO is like IFRS.

2021). Establishing robust standards is therefore essential to harnessing the potential of quantum technologies.

### *3.2 Research Design & Data Collection*

To identify relevant standard-setting bodies, a structured search was conducted in March 2026 using the search terms "quantum standards" and "quantum standard-setting organizations." The search combined an exploratory review of publicly available sources including institutional websites, standards catalogs, and official documentation describing quantum technology initiatives with AI-assisted screening using Perplexity, which was employed to surface candidate organization names and filter for relevance against the inclusion criteria below; all candidate entries were subsequently verified independently against primary institutional sources and all tables were cross referenced. A second independent pass through the ISO, NIST, ETSI, and ITU standards repositories was conducted to confirm coverage. Organizations were included in the analysis if they met at least one of the following criteria:

1. The organization publishes or develops formal standards related to quantum computing or quantum information technologies.
2. The organization coordinates standardization initiatives relevant to quantum technologies.
3. The organization plays a recognized role in international or national technology standard-setting that includes quantum-related activities.

For each organization identified, publicly available documentation from official websites and standards repositories was examined to determine institutional characteristics, governance structure, and stakeholder participation. Information on each organization was collected from primary institutional sources, including official documentation describing the organization's mandate, governance structure, and standards development processes. Data were extracted regarding:

- institutional authority
- organizational type
- enforcement mechanisms
- stakeholder participation

These attributes were selected because they are commonly used in accounting research examining regulatory and standard-setting institutions. In particular, prior accounting literature has emphasized how governance structures and enforcement mechanisms influence the adoption and credibility of standards. Organizations were then classified according to their institutional characteristics.

Table 2 summarizes the organizations identified and reports their authority, organizational type, enforcement mechanisms, and primary stakeholder groups. The resulting dataset exhibited in the following tables includes both global and regional standard-setting bodies as well as governmental organizations involved in technology standardization. These include international standards organizations, government agencies, and professional engineering bodies. The analysis focuses on the institutional structure of quantum standard setting, rather than on the technical content of individual standards. Specifically, the goal is to understand how different organizations contribute to the emerging governance framework surrounding quantum technologies. This institutional mapping approach allows the study to identify:

- the diversity of organizations involved in quantum standardization
- differences in governance and enforcement mechanisms
- the stakeholder groups participating in standard development

Understanding these institutional structures is relevant for accounting research because standards governing information technologies can influence the reliability, security, and auditability of systems used in financial reporting and auditing environments. The study documents the organizations participating in quantum standard-setting efforts and their institutional characteristics, establishing a foundation for analyzing how emerging quantum technologies may intersect with AIS and related assurance processes. Table 2 maps the principal quantum standard-setting organizations by authority, type, enforcement mechanism, and stakeholder base, providing the institutional foundation for the comparative analysis that follows.

| Organization | Authority | Type | Enforcement | Stakeholders |
| --- | --- | --- | --- | --- |
| International Organization for Standardization (ISO) | Global | Non-profit | Voluntary, certification, legal compliance | Manufacturers, developers, users, regulators, policymakers, educators, researchers, and consumers |

| National Institute of Standards & Technology (NIST) | U.S. federal government | Government agency | Voluntary, compliance with federal laws and regulations | U.S. government agencies, industry, academia |
|---|---|---|---|---|
| European Telecommunications Standards Institute (ETSI) | European Union | Non-profit | Voluntary, certification, legal compliance | European Union member states, industry |
| International Telecommunication Union (ITU) | United Nations | Intergovernmental organization | Voluntary, legal compliance with international treaties and agreements | Member states of the United Nations and ITU, industry |
| International Electrotechnical Commission (IEC) | Global | Non-profit | Voluntary, certification, legal compliance | Manufacturers, developers, users, regulators, policymakers, educators, researchers, and consumers |
| European Committee for Electrotechnical | European Union[2] | Non-profit | Voluntary, certification, legal compliance | European Union member states |

[2] CENELEC does not hold the status of an EU institution. Nonetheless, the standards it develops bear the designation of "EN" EU (and EEA) standards, attributable to the regulatory framework stipulated by EU Regulation 1025/2012.

| Standardization (CENELEC) | | | | |
|---|---|---|---|---|
| Institute of Electrical and Electronics Engineers (IEEE) | IEEE Standards Association | Non-profit | Voluntary | Engineers, scientists, researchers, educators, policymakers, and industry leaders |
| European Information Technologies Certification Institute Quantum Standards Group (EITC QSG) | European Union/Global[3] | Non-profit | Voluntary, certification, legal compliance | International experts in relevant fields who are interested in quantum technology and industry specifications and standards development |

**Table 2: Standard Setting Organizations Involved with Quantum Computing**

The standards produced by these organizations are generally voluntary and developed through consultative processes shaped by a mix of coercive influences from regulatory bodies, normative expectations within professional and technical communities, and mimetic tendencies among industry participants. These processes typically draw in manufacturers, developers, users, regulators, policymakers, researchers, and other stakeholders committed to the reliability and performance of quantum technologies. Through such collaboration, standard-setting organizations seek to establish commonly accepted technical frameworks that promote the development and deployment of quantum systems, often resulting in patterns of convergence and alignment across otherwise diverse contexts.

[3]EITCI does not hold the official status of an EU organization; however, it adheres to the guidelines set forth by the European Commission.

These organizations differ in their institutional authority, governance structures, enforcement mechanisms, and levels of stakeholder participation, which in turn influence the pressures they exert and the responses they elicit. Some function as global standard-setting bodies whose outputs achieve widespread adoption and isomorphism across jurisdictions, while others remain oriented toward regional or national contexts. In many cases, standards stay formally voluntary yet gain traction through mechanisms of diffusion, certification, or incorporation into regulatory frameworks, thereby reinforcing legitimacy and practical influence.

Despite variations in institutional configurations, coordination among these organizations is commonplace, reducing duplication and enhancing interoperability in emerging quantum technologies. Working groups within ISO and IEC, for example, align with IEEE initiatives to pinpoint standardization priorities in quantum computing, while industry consortia such as the Quantum Economic Development Consortium work in tandem with technical standards bodies to harmonize practices and accelerate convergence. Individual organizations further advance quantum standardization through their distinct institutional roles. The National Institute of Standards and Technology, for instance, contributes to measurement standards and cryptographic protocols for quantum information systems in response to both technical demands and broader regulatory expectations. International bodies like the International Telecommunication Union foster global coordination by bridging member states and industry actors, and Pan-European bodies such as the European Telecommunications Standards Institute (ETSI) whose membership spans EU and non-EU countries, including the United Kingdom post-Brexit develop specifications attuned to regional institutional environments.

The next section breaks down each standard-setting organization and its standards.

### *3.3 Analysis of Quantum Standards*

Building on the identification of principal standard-setting bodies, this section maps the quantum-related standards they have issued or are developing in quantum computing, communication, and cryptography. The analysis offers stakeholders from manufacturers and infrastructure providers to policymakers, auditors, and AIS designers actionable insight into how these institutional forces condition legitimacy, trust, and confidence in quantum products and services. This analysis emphasizes standards' role as institutional scripts that encode security baselines, interoperability requirements, ethical norms, and assurance expectations across jurisdictions. Systematically reviewing quantum standards, position papers, and related guidance from each organization, allows this section to exhibit how different bodies (e.g. US-oriented, EU-oriented, or global) construct distinct, yet overlapping, trajectories for quantum-safe infrastructures. This, in turn, illuminates how organizations engaged in accounting and financial reporting can align their quantum-related strategies with emerging field-level expectations, while remaining attentive to risks of symbolic compliance and decoupling between formal adherence to standards and actual implementation in AIS environments. Table 3 presents the ISO's active quantum-related standards and preparatory documents, illustrating how the organization's foundational vocabulary and cryptographic frameworks shape conformity expectations across accounting and critical-infrastructure environments.

**<Table 3, Appendix>**

ISO’s quantum-related work positions it as a key institutional platform for stabilizing language, expectations, and assurance practices around quantum technologies in information systems. Projects such as PQCRYPTO ICT-645622 illustrate ISO’s early engagement with post-quantum cryptography, signaling to regulators, vendors, and critical-infrastructure operators that long-term security against quantum attacks is a legitimate and urgent design consideration. Draft and preparatory documents like ISO/IEC 4879 (terminology for quantum computing in IT) and ISO/IEC TR 18157 (introductory technical report on quantum computing) play a foundational role in defining a shared vocabulary and conceptual map for the field

In the security domain, the ISO/IEC 23837 series (Parts 1 and 2 on security requirements, test methods, and evaluation for quantum key distribution) extends ISO’s long-standing role in information security management into the quantum realm. Through the codification of how QKD systems should be specified, tested, and evaluated, these standards create a template for conformity assessment and certification, establishing a certification baseline for vendors whose quantum-enabled communication systems meet internationally recognized security benchmarks. For financial infrastructures, this means that claims about “quantum-secure” communication links or key-management processes can, in principle, be tied to an ISO/IEC testable reference rather than purely proprietary assertions.

ISO’s “Computing Quantum Technologies Foresight” initiative complements these more formal standards by framing plausible trajectories, risks, and applications of quantum computing. For auditors, and regulators, ISO’s quantum vocabulary, QKD security standards, and foresight outputs collectively provide a global reference frame against which to interpret organizational strategies, risk disclosures, and control designs related to quantum computing and post-quantum security. Moving from this global vocabulary and reference frame to concrete algorithmic baselines, Table 4 catalogs NIST's active quantum and post-quantum cryptography (PQC) standards and initiatives, documenting how the agency's standardization work has shaped technical baselines and compliance expectations relevant to AIS environments.

**<Table 4, Appendix>**

NIST functions as the central institutional authority translating emergent post-quantum research into de facto global baselines for cryptographic practice. Its multi-year post-quantum cryptography (PQC) standardization process, culminating in the 2022 selection of candidate algorithms and ongoing work toward full standards, creates strong pressures: regulators, vendors, and infrastructure operators treat NIST-endorsed schemes as the reference point for "acceptable" post-quantum readiness. Beyond PQC, NIST's quantum communications research and testbeds support development of quantum-secure data links directly applicable to AIS infrastructure.

NIST's post-quantum and quantum-communications outputs function as both a blueprint and a benchmark. Audit, and financial market participants can rely on NIST standards and reports when specifying controls, assessing supplier offerings, or designing assurance procedures for cryptographic migrations. A key risk, however, is decoupling: firms may invoke NIST language in policies, disclosures, or "quantum-ready" marketing while only partially implementing the prescribed algorithms, key-management practices, and migration plans in their operational environments. This tension makes NIST's quantum-related work a reference point for evaluating the credibility of organizational claims about post-quantum cybersecurity and control over data. Table 5 documents ETSI's quantum-related standards, tracing how the organization's specialized work on Quantum-Safe Cryptography (QSC) and Quantum Key Distribution (QKD) has translated Pan-European regulatory and industrial priorities  notably, ETSI's membership extends beyond EU borders to include non-EU European countries including the United Kingdom post-Brexit into technical specifications with direct implications for network security and AIS-relevant infrastructures.

**<Table 5, Appendix>**

ETSI provides a comprehensive suite of Quantum-Safe Cryptography (QSC) and Quantum Key Distribution (QKD) standards. Its work on threat assessment, symmetric key sizes, algorithm selection, and migration strategies provides directly applicable guidance for AIS organizations. ETSI's QKD standards define a shared technical grammar for secure quantum-enabled networks, specifying device parameters, channel characteristics, software-defined networking interfaces, and protection profiles (e.g., REST-based key-delivery APIs). This specificity promotes interoperability across network operators and vendors supporting vendor interoperability and AIS security conformance assessments.

Firms may reference ETSI QSC and QKD documents in policies or assurance reports while only partially realizing the underlying architectures and controls in practice, a decoupling risk that AIS auditors should assess. Table 6 maps the ITU's quantum-related standards and initiatives, illustrating how the organization's global coordination mandate has translated into technical guidance for quantum communication and cryptography across member states.

**<Table 6, Appendix>**

The ITU's standardization pillar (ITU-T) develops recommendations on quantum key distribution (QKD) network architectures, functional requirements, interworking, and key management, providing globally applicable technical baselines for quantum-secure network infrastructure. ITU-T outputs are relevant as reference models for designing or auditing quantum-enabled communication links that may carry financial and AIS-related traffic.

At the same time, the voluntary character of ITU standards and the diversity of national regulatory frameworks mean that implementation remains uneven, creating scope for decoupling between formal adherence to ITU-aligned language in policy or technical documentation and the depth of actual conformity in deployed systems. Table 7 maps the IEC's quantum-related standards and joint ISO/IEC initiatives, showing how IEC's conformity assessment role shapes security expectations for manufacturers and AIS operators.

**<Table 7, Appendix>**

IEC and joint ISO/IEC standards provide internationally recognized conformity assessment frameworks relevant to quantum computing,

communication, and cryptography. These standards establish testable benchmarks that organizations can reference when evaluating vendor claims and securing supply chains against quantum threats.

Of particular relevance to AIS, ISO/IEC 9594-11:2020 specifies a general "wrapper protocol" providing authentication, integrity, and confidentiality across multiple protocols with explicit guidance on cryptographic algorithm migration to quantum-safe algorithms. This directly bridges today's PKI-based AIS infrastructures and future quantum-safe suites. AIS auditors should verify that references to IEC/ISO/IEC standards in policies and disclosures reflect actual implementation, not merely symbolic adoption. Table 8 catalogs CENELEC's quantum-related standardization activities, showing how the organization bridges cutting-edge quantum research and EU industrial policy to codify deployment expectations for European firms operating in AIS-relevant sectors.

**<Table 8, Appendix>**

CENELEC's Joint Technical Committee 22 (JTC 22) coordinates European quantum standards across computing, communication, cryptography, and sensing directly relevant to AIS security and infrastructure planning. The JTC 22 Standardization Roadmap identifies where European quantum standards will emerge, providing practitioners with advance visibility into evolving compliance expectations. The recently published CEN/CLC/TR 18202:2025 layer model for quantum computing is particularly significant. It abstracts hardware-software layers across multiple qubit technologies, establishing the foundation for interface standards that will govern how quantum co-processors integrate into ERP and AIS environments.

CENELEC's coordination with IEC under the Dresden and Frankfurt Agreements produces harmonized European and international standards, ensuring that quantum-related controls and architectures for AIS are consistent with both EU regulatory requirements (e.g., cybersecurity and data protection) and globally recognized technical baselines. Table 9 catalogs IEEE's quantum-related standards and initiatives, illustrating how IEEE's market- and innovation-driven standards define quantum architecture baselines, security practices, and performance benchmarks relevant to AIS design and audit.

**<Table 9, Appendix>**

IEEE standards are widely treated as de facto benchmarks for quantum architectures, security practices, and performance metrics across the technology industry. This makes IEEE a key reference point when evaluating quantum-related vendor claims, system designs, and audit evidence.

Architecture and integration efforts (e.g. standards for quantum and programmable simulators, hybrid quantum, classical systems, and quantum performance metrics and benchmarking) define common reference models and evaluation criteria that organizations can adopt when designing or procuring quantum capabilities. Post-quantum network security, cryptography migration, and quantum algorithm design guidelines provide concrete implementation templates for hardening AIS against quantum-enabled threats. Semantic and management standards, such as terminology definitions and YANG models for software-defined quantum communication, support interoperability and coordination in software-defined environments.

Since IEEE standards are voluntary, adoption is uneven and decoupling is possible, organizations may cite IEEE quantum standards in disclosures without fully embedding the underlying controls. For assurance, this means IEEE standards serve both as practical guidance on quantum architectures and security, and as a benchmark for evaluating whether reported compliance reflects actual quantum-related controls. Table 10 catalogs the EITCI Quantum Standards Group's (QSG) quantum-related standards and certifications, documenting how this European industry-based body translates academic and regulatory expectations into practitioner-facing credentials and technical benchmarks for quantum-ready AIS.

**<Table 10, Appendix>**

The EITCI Quantum Standards Group (QSG) is a practitioner-oriented body developing technical specifications for QKD, QRNG, and quantum encryption protocols, with outputs that feed into formal SDOs (ETSI, ISO/IEC, ITU). Its work is relevant to AIS in two respects: QSG specifications define what constitutes a conformant quantum-secure communication implementation, providing AIS auditors with technical benchmarks for evaluating vendor claims; and EITCI certifications may appear in staff credentials and vendor disclosures, warranting evaluation as part of AIS assurance procedures.

### *3.4 Analysis of the path to adopting Quantum Standards*

Analysis of the surveyed standards and their issuing bodies reveals a clear institutional pattern: the majority of active standard-setting organizations are concentrated in the United States and Europe, with most substantive activity emerging in the 2020s and limited precursors in the 2010s. This temporal and geographic clustering mirrors the maturation of quantum technologies from experimental constructs to commercially deployable systems and illustrates how coercive, normative, and mimetic pressures converge to accelerate isomorphism in post-quantum cryptography. When organizations contemplate migration to quantum-safe standards, they confront a multifaceted set of institutional influences that shape adoption decisions. Standard-setting bodies particularly NIST offer explicit guidance that integrates these pressures, emphasising the following considerations:

- **Security imperatives:** The primary driver remains protection against quantum-enabled threats. Standards organisations stress selection of algorithms that deliver robust, future-proof security guarantees, thereby satisfying coercive regulatory demands and normative expectations for data integrity.
- **Resource and cost implications:** Adoption entails significant financial outlays for hardware/software upgrades, cryptographic library migrations, and system re-engineering. Organizations must weigh these against resource-dependence constraints to avoid decoupling - i.e., symbolic compliance without substantive change.
- **Knowledge and expertise requirements:** Quantum-safe cryptography demands specialised capabilities. Firms respond to normative professional pressures by investing in staff training or external expertise, aligning with the institutional logics of technical communities.
- **Interoperability and compatibility:** Heterogeneous legacy environments necessitate algorithms that preserve operational continuity. Mimetic pressures encourage alignment with peer-adopted solutions to reduce integration friction and promote system-wide convergence.
- **Performance trade-offs:** New algorithms may alter speed and efficiency profiles. Organisations assess these characteristics to ensure acceptable operational outcomes while maintaining legitimacy in stakeholder eyes.
- **Long-term viability:** Preference is given to standards subjected to prolonged scrutiny (e.g., NIST's multi-round evaluation processes), supporting sustained isomorphism amid rapid technological evolution.
- **Regulatory compliance:** Industry- and jurisdiction-specific mandates constitute coercive institutional forces; early alignment minimises legal exposure and reinforces organisational legitimacy.

- **Validation and testing:** Rigorous pre-production testing and iterative validation are prescribed to confirm correctness, security, and interoperability practices that embed normative quality benchmarks.
- **Transition governance:** A structured migration plan with clear timelines, milestones, and contingencies mitigates disruption risks and facilitates coordinated organisational responses to institutional change.
- **Risk assessment:** Comprehensive evaluation of vulnerabilities and interdependencies enables targeted mitigation, ensuring that adoption enhances rather than erodes institutional legitimacy.

Systematic attention to these factors positions organisations to execute a transition that safeguards data and communications while minimising operational friction. Standard-setting bodies supply the institutional scaffolding that underpins a secure, convergent quantum-safe cryptography, providing the resources, evaluation frameworks, and consensus-driven benchmarks on which that transition depends. The ultimate costs of development and implementation will scale with quantum-computing progress and the pace of standard diffusion; proactive alignment with these institutional pressures therefore represents both a defensive necessity and a source of competitive legitimacy in an evolving regulatory and market environment.

### *3.5 Differences & Similarities between quantum standards*

The observed differences between US and European approaches to quantum standards do not arise purely from technical preference. The differences are rooted in deeper institutional and regulatory philosophies that have historically shaped how each jurisdiction governs.

Monopolies have long been a concern in the United States and the European Union, albeit with distinct emphases attributed to their historical and philosophical differences (Bradford, 2020). In the US, the prevailing viewpoint is that monopolies are detrimental to consumers. We can trace this perspective back to historical antitrust legislation and the philosophical underpinnings of American capitalism. The United States has historically grappled with monopolistic practices since the late 19th century when industrial giants like Standard Oil and the American Tobacco Company dominated their respective markets (Sawyer, 2019). The emergence of these monopolies prompted lawmakers to enact antitrust laws, notably the Sherman Antitrust Act of 1890, which aimed to preserve competition and protect consumers from unfair business practices (Sherman Anti-Trust Act, 1890). This historical context has contributed to the prevailing American belief that monopolies can stifle innovation, limit consumer choice, and drive up prices. Philosophically, the US is strongly committed to free-market capitalism, rooted in the ideals of competition and individual entrepreneurship. The belief in the power of competition to drive economic growth and innovation has been a cornerstone of American economic thought (Kovacic, 2009). Consequently, the skepticism toward monopolies aligns with the overarching philosophy that a competitive marketplace benefits society.

On the other side of the Atlantic, the European Union's competition regime reflects a systematic commitment to preserving effective competition and preventing the creation or maintenance of market power that would harm consumers. As articulated in the Commission's Guidelines on the application of Article 81(3), the objective of EU competition law is to protect competition on the market as a means of enhancing consumer welfare and ensuring an efficient allocation of resources. In this framework, monopolistic or dominant positions are scrutinized insofar as they restrict rivalry, reduce innovation, or worsen price, output, quality, or variety for consumers (European Commission, 2004).

Thus institutional differences help explain why European quantum standards tend to emphasize cross-member-state coordination, common terminology, and alignment with broader frameworks such as data protection and industrial policy. The resulting standards reinforce isomorphism around shared vocabulary, minimum security baselines, and interoperability requirements that can be audited and enforced within the EU's legal architecture. In the US context, standard-setting is more frequently driven by industry consortia, research alliances, and federal agencies seeking to preserve technological leadership. American standards often function as non-binding coordination devices that facilitate experimentation across application domains, with organizations imitating perceived technological leaders to maintain legitimacy in the eyes of investors, clients, and regulators. EU-oriented and US quantum standards are shaped by different configurations of coercive, normative, and mimetic pressures rather than purely technical considerations. EU-aligned bodies such as CENELEC and EITCI operate within a supranational regulatory environment that prioritizes harmonization, data protection, and social welfare objectives, creating strong coercive and normative expectations around privacy, cybersecurity, and ethical use of technology. By contrast, US-oriented processes linked to organizations such as IEEE are more tightly coupled to market logics, venture funding, and industry coalitions, where mimetic and competitive pressures to innovate dominate and formal coercive mandates remain more fragmented.

Competition policy traditions also feed into these institutional logics. EU competition and antitrust regimes reinforce an emphasis on preventing dominance and ensuring access, which supports standards designed to avoid lock-in and to protect smaller actors within the quantum ecosystem. In practice, this can translate into stronger attention to interoperability, fair access to infrastructures, and constraints on proprietary control over critical interfaces. In the US, a more permissive stance toward market concentration, combined with a strong ideological commitment to entrepreneurship and innovation, legitimizes standards processes that may be more closely aligned with leading firms' architectures and de facto platforms. This can accelerate diffusion of particular technical solutions but also raises the risk that standards entrench specific commercial ecosystems.

For AIS and financial reporting, these institutional contrasts imply different forms of legitimacy attached to standards claims. European-aligned quantum standards can serve as signals of regulatory and normative conformity showing that firms' cryptographic and quantum-related controls are consistent with societal expectations around privacy, ethics, and harmonization. US-aligned standards, in turn, primarily convey pragmatic and performance legitimacy, indicating responsiveness to cutting-edge practice and industry consensus. In both cases, however, institutional theory highlights the possibility of decoupling: organizations may publicly reference European or US quantum standards to secure legitimacy with key audiences while only partially embedding the associated controls and governance mechanisms in their AIS. Understanding these similarities and differences is therefore essential for assessing how quantum standards will actually shape security, comparability, and assurance in AIS.

### *3.6 Quo vadis - Limitations of Standards*

The contrast between EU-oriented (and, where relevant, broader Pan-European) and US approaches illustrates how different institutional environments privilege distinct logics. In the EU, standard setters tend to embed quantum standards within broader societal imperatives, producing strong normative expectations around alignment with public-interest goals and common terminology. This orientation supports field-level isomorphism around shared concepts and minimum security baselines, but may constrain flexibility and responsiveness to fast-moving technological niches, especially where market experimentation could reveal superior solutions.

In the US, by contrast, quantum standards are more tightly coupled to market-driven innovation and industry engagement, with federal agencies and industry consortia emphasizing guidance that facilitates experimentation, interoperability, and sector-specific adaptation. This institutional configuration encourages mimetic and competitive isomorphism while coercive pressures remain weaker and more fragmented across jurisdictions. The result is a dynamic but potentially uneven standards environment, where interoperability and performance are prioritized, yet privacy, ethics, and harmonized assurance practices risk being under-specified.

For financial reporting, these differences matter because they shape the kinds of legitimacy signals standards can provide. EU-oriented standards can confer normative and regulatory legitimacy by demonstrating alignment with societal and regulatory expectations, whereas US-style standards tend to foreground pragmatic legitimacy showing responsiveness to client demands, performance, and innovation. In both settings, discrepancies between claimed adherence to quantum standards and the underlying technical reality create a risk of decoupling, with direct implications for auditability and investor confidence.

Institutional theory also highlights how the voluntary status of most quantum standards limits coercive pressure and increases the scope for symbolic adoption. In the absence of binding legal mandates, organizations face heterogeneous incentive structures: some adopt quantum-safe practices to pre-empt regulation, satisfy powerful stakeholders, or signal technological sophistication; others delay or adopt only superficial elements. This produces uneven isomorphism, where leading firms and critical infrastructures move first while smaller or less visible entities lag, creating systemic vulnerabilities despite the existence of standards. Moreover, the advances in quantum hardware, algorithms, and cryptographic techniques outstrips the cycle time of formal standard-setting processes. When formal standards are perceived as outdated or incomplete, organizations increasingly look to peers, key vendors, and professional networks for cues about "good practice," engaging in mimetic, interorganizational imitation to reduce uncertainty and maintain legitimacy (Haunschild & Miner, 1997). As a result, standards may serve more as retrospective codifications of emerging practice than as stable, prospective guides. This mirrors the broader challenge identified in the tax digitalization literature, where the rapid adoption of AI and blockchain technologies in tax systems has outpaced the development of adequate regulation (Belahouaoui & Attak, 2024).

## IV. Implications & Challenges

### *4.1 Implications of Quantum Standards for AIS*

The analysis of standard-setting bodies highlights that current efforts are fragmented, largely voluntary, and concentrated in the development of post-quantum cryptography and secure communication protocols. As a result, the transition to quantum-safe infrastructures is likely to be uneven and institutionally mediated.

Quantum vulnerability is most acute at the transaction processing layer. ERP systems and transaction processing components rely on public-key encryption schemes that quantum algorithms can compromise. If such vulnerabilities materialize, the integrity and verifiability of financial records could be undermined, with direct consequences for compliance with financial reporting and internal control requirements (Mojtahedi & Zhou, 2024). Emerging quantum standards related to post-quantum cryptography provide a pathway for mitigating these risks; however, as shown in the preceding analysis, these standards remain under development and are primarily voluntary, creating uncertainty for organizations seeking to align AIS transaction infrastructure with best practices.

At the general ledger and consolidation layer, quantum threats introduce risks to the cryptographic foundations that underpin data integrity across period-end reporting and multi-entity consolidations. Blockchain-based accounting applications are particularly exposed, as the erosion of cryptographic security challenges claims of immutability and auditability that these systems depend on for evidentiary value (Khodaiemehr et al., 2026). The voluntary and evolving nature of current quantum standards may also produce inconsistencies in implementation across jurisdictions and firms, with direct implications for the comparability of consolidated financial statements and the reliability of XBRL-enabled reporting and sustainability disclosures.

Audit analytics platforms depend on the integrity of digital audit trails and the reliability of data pipelines that feed sampling, anomaly detection, and continuous monitoring routines. Quantum-induced compromise of the underlying cryptographic infrastructure would degrade the evidentiary quality of data produced by these platforms, complicating the formation of audit opinions and challenging the reliability of automated assurance procedures. Upgrading these systems to quantum-resistant architectures involves significant cost, complexity, and timing uncertainty in the absence of fully stabilized standards. This places standard setters, regulators, and professional bodies in a critical role, as they translate technical developments into actionable guidance for assurance practice. The adoption of quantum-resistant audit infrastructure will not occur as a discrete technological shift, but rather as a gradual and uneven process embedded within existing standard-setting and reporting regimes.

Quantum standards must be reconciled with existing regulatory frameworks for data protection, financial reporting, and cybersecurity. Quantum standards do not displace existing frameworks for internal control, audit evidence, and disclosure; they sit alongside them, creating a more complex compliance environment than either set of requirements alone would generate. For SOC reporting and cyber-assurance engagements, this raises practical questions about which quantum-related controls are sufficient to satisfy existing regulatory duties, and how auditors and regulators weigh claims of quantum-standard compliance when forming opinions on control effectiveness and financial statement reliability. The adoption of quantum-resistant assurance practices is likely to be shaped by coercive pressures such as regulatory cybersecurity requirements, normative pressures from professional guidance and audit standards bodies, and mimetic behavior among firms responding to peer adoption under uncertainty, a dynamic not unlike the tax digitalization context, where rapid adoption of AI and blockchain technologies in tax systems has outpaced the development of adequate regulation (Belahouaoui & Attak, 2024). Institutionally, this implies a field in transition, where quantum standards simultaneously function as vehicles for institutional change (by introducing new expectations around quantum-safe cryptography, interoperability, and assurance) and as potential sites of decoupling (where compliance is asserted but only partially implemented). Understanding these dynamics is essential for evaluating how quantum standards will actually shape the security, reliability, and legitimacy of AIS over time.

The voluntary and evolving nature of current quantum standards may lead to inconsistencies in implementation across jurisdictions and firms, with potential implications for comparability and auditability. There is limited empirical evidence on how key actors (preparers, auditors, regulators, and technology providers) perceive and respond to quantum-related risks, constraining the ability to assess real-world impacts on accounting practice.

These challenges point to a need for research that is closely aligned with accounting standard-setting and policy concerns. In particular, future studies should examine how quantum-resistant standards influence AIS design and internal controls, how they affect the assurance of financial and non-financial disclosures, and how institutional environments shape their adoption and enforcement. Such work would extend the current analysis by linking the emerging quantum standards landscape to observable outcomes in accounting and auditing practice.

### *4.2 Future Research & Limitations*

A central limitation of the current literature is the lack of accounting-focused, enterprise-level evidence on the adoption of quantum technologies. While existing research highlights applications of quantum computing in areas such as logistics, chemistry, and risk analysis, there is little insight into how these technologies can be operationalized within AIS or how they may affect financial reporting and assurance outcomes. This limits the ability of accounting researchers to move beyond conceptual analysis and assess organizational implications empirically.

In the absence of observable firm-level implementations and publicly available data, the literature remains largely theoretical, focusing on emerging standards, conceptual frameworks, and system design considerations. Although adjacent fields such as FinTech offer examples of advanced simulation and hybrid modeling approaches, their relevance to AIS, internal controls, and reporting processes is not yet well established.

Moreover, existing simulation-based studies are primarily technical in nature and do not address accounting-specific concerns. For instance, simulations of quantum architectures demonstrate computational feasibility but provide limited insight into AIS design, auditability, or compliance with reporting standards. As a result, their contribution to accounting research and standard-setting remains indirect.

Future research should prioritize the development of empirically grounded, accounting-relevant approaches. This includes modeling the impact of quantum-resistant standards on AIS architecture, examining how cryptographic transitions affect internal controls and audit processes, and assessing implications for the reliability and comparability of financial and non-financial disclosures. As early adoption emerges and data availability improves, opportunities for archival, field-based, and mixed-method research will become more feasible.

## V. Conclusion

This study has mapped the emerging landscape of quantum computing standards and demonstrated their relevance for the design and governance of AIS, particularly in the domain of cybersecurity. By surveying major standard-setting bodies in the United States and Europe, the analysis shows how voluntary quantum standards, post-quantum cryptography projects, and QKD/QRNG initiatives collectively shape expectations about interoperability, security, and the future trajectory of quantum technologies.

Viewed through an institutional lens, quantum standards function as coercive, normative, and mimetic pressures that encourage organizations to converge on quantum-safe architectures, even while large-scale fault-tolerant computing remains nascent. Legitimacy and political-economic perspectives clarify why regulators, professional bodies, vendors, and preparers contest the content and timing of these standards, and how differences between European, US, and consortium-based approaches reflect broader societal priorities, regulatory philosophies, and market logics. For AIS, this implies that quantum cybersecurity is a standards-driven institutional change process that will be scrutinized by auditors, regulators, and capital markets.

For practice, the findings underscore that accounting and assurance functions should monitor quantum standard-setting processes, plan staged migration paths to quantum-resilient cryptography, and evaluate the cost, performance, and ESG implications of candidate solutions in line with emerging guidance from standard-setting bodies. For research, the paper opens an agenda at the intersection of AIS, cybersecurity, and standard setting: future work can examine how different institutional environments affect quantum-standards adoption, how disclosures about quantum readiness influence perceived legitimacy, and how evolving quantum standards reshape the auditability and governance of digital financial infrastructures. Just as the reliability of financial records has always depended on the strength of the controls surrounding them, the integrity of quantum-era AIS will depend on whether the voluntary standards governing those choices are substantive enough to be trusted by auditors, regulators, and capital markets.

**Declarations**

***Ethical Approval and Consent to participate***

This article does not involve any studies on human participants or animals conducted by the author. No ethical approval was required for this theoretical research paper that examines quantum standards and accounting information systems through conceptual analysis and literature review.

***Consent for publication***

Not applicable. This manuscript does not contain any individual person's data requiring consent for publication.

***Availability of supporting data***

The datasets used and/or analyzed during the current study are publicly available. All data supporting the conclusions of this article are included within the manuscript and are based on publicly accessible literature, standards documentation, and theoretical frameworks in quantum computing and accounting information systems. No additional raw data were generated or analyzed during this study.

***Competing interests***

The author declares no competing interests. There are no financial or non-financial interests that are directly or indirectly related to the work submitted for publication.

***Authors' contributions***

As the sole author, I conceived and designed the study, conducted the literature review and analysis, developed the theoretical framework, and drafted the manuscript. I take full responsibility for the content and conclusions presented in this work.

***Funding***

No funding was received for conducting this study. This research received no specific grant from any funding agency in the public, commercial, or not-for-profit sectors.

**Citations:**

---

## Appendix:

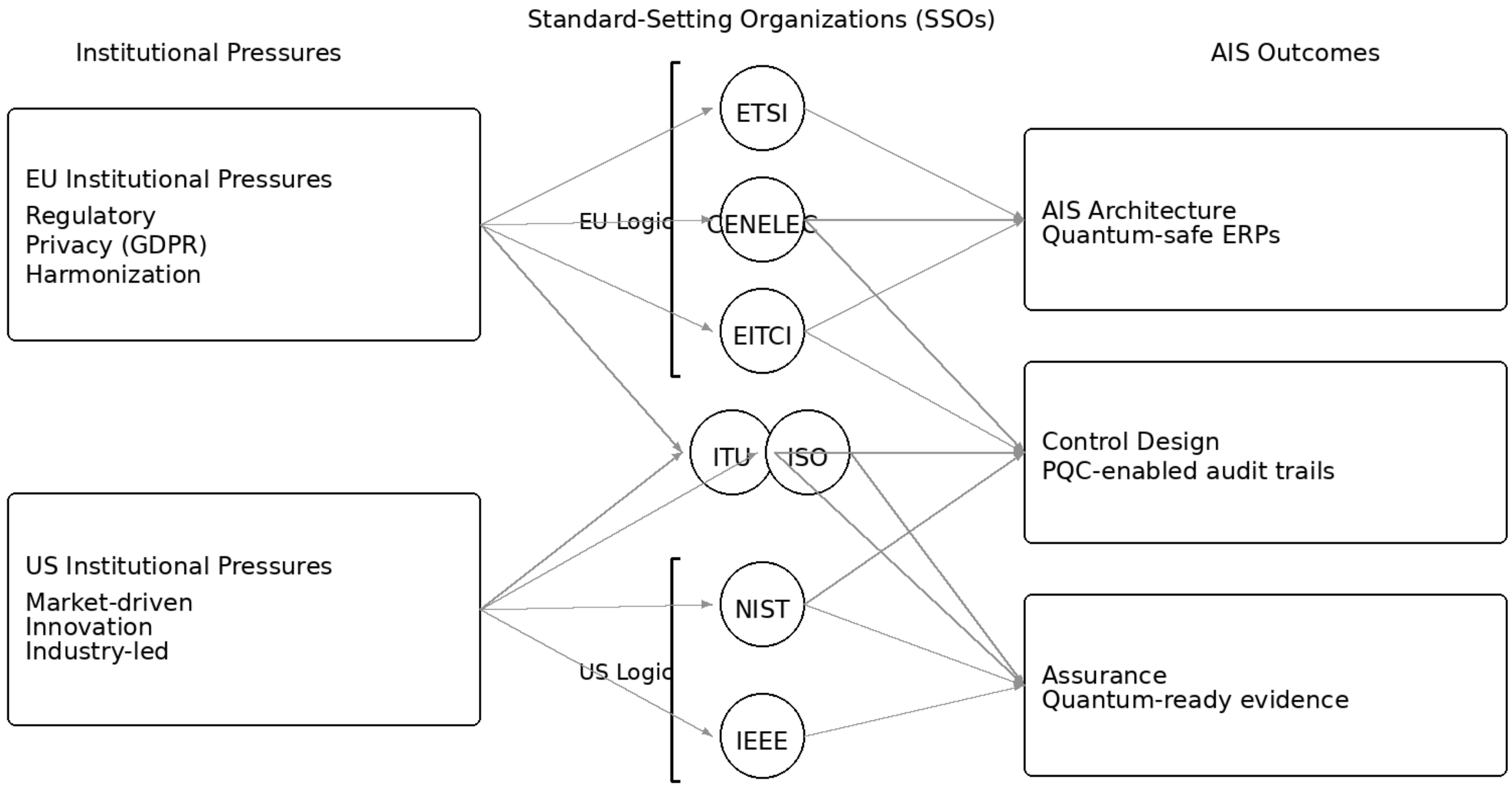

**Graphical Abstract**

| Standard | Title/Description | Status | Notes |
|---|---|---|---|
| PQCRYPTO ICT-645622 | Post-quantum cryptography for long-term security | Completed (2015-2018) | EU Horizon 2020 project; informed ISO PQC efforts; |

| Standard | Title/Description | Status | Notes |
| --- | --- | --- | --- |
| | | | no active updates.c |
| ISO/IEC 4879 | Information technology — Quantum computing — Terminology and vocabulary | Published (2024) | Finalized post-2023 voting; defines core terms for interoperability. |
| ISO/IEC TR 18157 | Information technology — Introduction to quantum computing | Preparatory/enquiry stage | Ongoing since 2022; technical report under development. |
| ISO/IEC 23837-1 | Information security — Security requirements, test and evaluation methods for quantum key distribution — Part 1: Requirements | Published (2023) | Framework for QKD evaluation per ISO/IEC 15408. |
| ISO/IEC 23837-2 | Information security — Security requirements, test and evaluation methods for quantum key distribution — Part 2: Evaluation and testing methods | Published (2023) | Complements Part 1; focuses on testing protocols. |
| ISO/IEC JTC 3 | Quantum technologies (new committee) | Active (since 2024) | Joint IEC/ISO; covers computing, sensing, communications; U.S. TAG recruiting. |
| Computing Quantum Technologies Foresight | Strategic foresight on quantum computing | Ongoing | ISO foresight page; guides standardization priorities. |

**Table 3: International Organization for Standardization (ISO)**

| Standard | Title/Description | Status | Notes |
|---|---|---|---|
| PQC Standardization | Post-Quantum Cryptography Selected Algorithms | Ongoing | Core for AIS quantum-safe key exchange/signatures; migrate now. |
| NIST IR 8545 | Status Report on Fourth Round PQC | Published Mar 2025 | Details HQC selection as ML-KEM backup. |
| NIST IR 8413 | Third Round PQC Status Report | Published 2022 | Round 3 outcomes leading to FIPS. |
| NIST CSWP 15 | Getting Ready for Post-Quantum Cryptography | Published Apr 2021 | Adoption challenges for systems like AIS. |
| NIST IR 8105 | Report on Post-Quantum Cryptography | Published Apr 2016 | Early PQC assessment. |
| Glossary: PQC | Post-Quantum Cryptography Definition | Updated Jul 2023 | Standardized terminology. |
| Quantum Communications | Quantum network/comms research | Active | QKD prototypes for secure AIS data links. |
| Quantum Networks at NIST | Quantum network testbeds | Active | Interoperable quantum-secure networks. |
| Post-Quantum Cryptography: Digital Signatures | PQC-Dig-Sig Standardization Examples | Updated Jul 2023 | Implementation files for ML-DSA/SLH-DSA. |
| Cryptographic Standards in Post-Quantum Era | Journal Article on PQC Transition | Published Nov 2022 | AIS governance implications. |

| Standard | Title/Description | Status | Notes |
|---|---|---|---|
| Workshop: Cybersecurity in Post-Quantum World | 2015 PQC Workshop | Archived Jun 2020 | Historical; initiated standardization. |

**Table 4: National Institute of Standards Technology (NIST)**

| Standard | Title/Description | Status | Notes |
|---|---|---|---|
| GR QSC 001 | Quantum-Safe Cryptography; Quantum-safe algorithmic framework | Jul 2016 (v1.1.1) | Framework overview; stable. |
| GR QSC 003 | Quantum-Safe Cryptography; Case studies and deployment scenarios | Feb 2017 | Deployment guidance. |
| GR QSC 004 | Quantum-Safe threat assessment | Mar 2017 | Threat modeling. |
| GR QSC 006 | Limits to quantum computing applied to symmetric key sizes | Feb 2017 | Symmetric crypto sizing. |
| TR 103 570 | Quantum-Safe Key Exchanges | Oct 2017 | Early KEM designs. |
| EG 203 310 | Quantum computing impact on ICT; Business continuity recommendations | Jun 2016 | Migration recs for AIS-like systems. |
| TS 103 744 | Quantum-safe Hybrid Key Exchanges | Dec 2020 | Hybrid PQC-classical. |

| Standard | Title/Description | Status | Notes |
|---|---|---|---|
| TR 103 617 | Quantum-Safe Virtual Private Networks | Sep 2018 | VPN migration. |
| TR 103 619 | Migration strategies to quantum-safe schemes | Jul 2020 | Roadmap. |
| TR 103 616 | Quantum-Safe Signatures | Sep 2021 | Sig schemes. |
| TR 103 823 | Quantum-Safe Public-Key Encryption and KEM | Oct 2021 (v1.1.2) | Aligns with NIST PQC. |
| TR 103 618 | Quantum-Safe Identity-Based Encryption | Dec 2019 | IBE variants. |
| TR 103 949 | QSC Migration; ITS/C-ITS study | May 2023 | Automotive focus. |
| New: Covercrypt (KEMAC) | Key Encapsulation with Access Control | Mar 2025 | Quantum-safe encryption std. |
| GS QKD 002 | QKD Use Cases | Jun 2010 | Applications incl. secure comms. |
| GS QKD 004 | QKD Application Interface | Dec 2010 | API specs. |
| GS QKD 005 | QKD Security Proofs | Dec 2010 | Proof frameworks. |
| GS QKD 008 | QKD Module Security Specification | Dec 2010 | Hardware security. |
| GS QKD 011 | Component characterization: | May 2016 | Testing params. |

| Standard | Title/Description | Status | Notes |
|---|---|---|---|
| | Optical components | | |
| GS QKD 012 | Device and channel params for QKD deployment | Feb 2019 | Deployment guidelines. |
| GR QKD 003 v2 | QKD Components and Internal Interfaces | Mar 2018 | Updated interfaces. |
| GR QKD 007 | QKD Vocabulary | Dec 2018 | Terminology. |
| GS QKD 014 | REST-based key delivery API | Feb 2019 | SDN integration. |
| GS QKD 015 v2 | Control Interface for SDN | Apr 2022 | SDN control. |
| GS QKD 018 | Orchestration Interface for SDN | Apr 2022 | Network orchestration. |
| GS QKD 016 | Common Criteria Protection Profile | Apr 2023; validated 2024 | First certifiable QKD PP. |

**Table 5: European Telecommunications Standards Institute (ETSI**

| Standard | Title/Description | Status | Notes |
|---|---|---|---|
| ITU-T Y.3800 | Overview on networks supporting QKD | 10/2019 | High-level overview of QKD networks (QKDN), baseline for subsequent Y. |
| ITU-T Y.3801 | Functional requirements for QKD networks | 10/2020 | Defines functional requirements for QKDN (nodes, links, key management). |
| ITU-T Y.3802 | QKD networks – Functional architecture | 10/2020 | Specifies reference architecture and key management entities |

| Standard | Title/Description | Status | Notes |
|---|---|---|---|
| | | | (QKDN controller/manager). |
| ITU-T Y.3813 | QKD network interworking – Functional requirements | 09/2024 | Interworking/QKDN-i requirements for multi-domain QKD networks. |
| ITU-T Y.3814 | QKD networks – Functional requirements and architecture for ML enablement | 01/2023 | Adds machine-learning enabled QKDN functions and architecture extensions. |
| ITU-T X.1714 | Key combination and confidential key supply for QKD networks | 10/2020 | Describes key combination and secure key supply across QKD networks. |
| ITU-T X.1715 | Security requirements and measures for integration of QKD networks and secure storage | 2022 | Complements X.1714 with integration and storage security controls. |
| ITU-T Y Suppl. 74 | Y.3800-series – Standardization roadmap on QKD networks | 03/2023 (approved; often cited as 2023/2024 roadmap) | Roadmap summarizing >40 QKD work items and mapping SDO landscape. |
| ITU-T Y Suppl. 75 | Y.3000-series – Quantum-enabled future networks | 03/2023 | Foresight on quantum-enabled future networks, identifies QKD priority research directions 2023–2030. |

**Table 6: International Telecommunication Union (ITU)**

| Standard | Title/Description | Status | Notes |
|---|---|---|---|
| IEC White Paper QIT:2021 | Quantum information technology | Published 2021-10-19 | Strategic overview of quantum information technology (QIT), mapping technology readiness, market readiness, and standardization readiness, including gaps and coordination needs across quantum computing, communications, sensing, and metrology.This supports your institutional argument that standardization is lagging underlying research and requires coordinated governance. |
| ISO/IEC 9594-11:2020 (≙ ITU-T X.510) | Information technology – Open Systems Interconnection – The Directory: Protocol specifications for secure operations | Published 2020-12; status “to be revised” (stage 90.92) | Specifies a general “wrapper protocol” providing authentication, integrity, and confidentiality for other protocols, with explicit guidance on cryptographic algorithm migration and auxiliary algorithms. The scope statement notes that the wrapper allows smooth migration to stronger, quantum-safe algorithms, so protocols protected by it can evolve without redesigning higher-level logic.This is a strong, concrete bridge between today’s PKI-based AIS infrastructures and future quantum-safe suites. |
| ISO/IEC JTC 3 | Standardization in the field of | Approved 2024; | New joint ISO/IEC committee coordinating |

| Standard | Title/Description | Status | Notes |
|---|---|---|---|
| (Quantum Technologies) | quantum technologies (computing, simulation, communications, metrology, etc.) | Secretariat: BSI; Chair: Republic of Korea | quantum technologies standards across quantum computing, communications (including QKD), sources, detectors, and metrology.It provides the institutional umbrella under which future IEC/ISO quantum standards relevant to AIS (e.g., secure quantum interfaces, quantum device characterization) will be harmonized. |

**Table 7: International Electrotechnical Commission (IEC)**

| Standard | Title/Description | Status | Notes |
|---|---|---|---|
| Quantum Technologies topic page | Public landing page summarizing European quantum standardization activities | Continuously updated (JTC 22 established 2023) | Describes CEN-CENELEC/JTC 22 scope across quantum computing, communication/cryptography, sensing/metrology, and enabling technologies, and its coordination with ETSI, EURAMET, EuroQCI, EuroHPC, and ISO/IEC JTC 3. Signals EU policy coupling via the Quantum Europe Strategy and frames standards as tools for interoperability and commercialization. |
| CEN-CENELEC/JTC 22 Standardization Roadmap on Quantum Technologies | European standardization roadmap for quantum technologies | Release 1.0: Mar 2023; Release 1.1: 2025; Release 2.0 expected end-2026 | Analyses market needs and timelines for quantum standards in computing, communication (including QKD and quantum-safe cryptography), sensing, and enabling technologies, |

| Standard | Title/Description | Status | Notes |
|---|---|---|---|
| | | | and is used to coordinate with IEC/ISO JTC 3 and ETSI TC .Provides an institutional blueprint for where European quantum standards will emerge, shaping long-term expectations for AIS-relevant security and infrastructure. |
| CEN/CLC/TR 18202:2025 | "Layer model of Quantum Computing" Technical Report | Published 2025-09-03 | First concrete JTC 22 deliverable; presents a high-level functional layer model for universal gate-based quantum computing across platforms (transmon, spin-qubit, ion-trap, neutral-atom, etc.).By abstracting hardware-software layers, it supports future interface and interoperability standards, which will matter for how quantum co-processors integrate into ERP/AIS and audit analytics stacks. |
| Putting Science Into Standards (PSIS) – "Making Quantum Technology Ready for Industry" | 6th PSIS workshop, Brussels, 28–29 March 2019 | Completed 2019 | Brought together researchers, industry, and standardizers to identify early standardization priorities and gap areas for quantum technologies. 1 Acts as a precursor to JTC 22 and the roadmap, illustrating how scientific developments are translated into candidate |

| Standard | Title/Description | Status | Notes |
|---|---|---|---|
| | | | standards topics—an example of the institutional pipeline from research to standardization that your paper theorizes. |

**Table 8: European Committee for Electrotechnical Standardization (CENELEC)**

| Standard | Title/Description | Status | Notes |
|---|---|---|---|
| P3329 | Standard for Quantum Computing Energy Efficiency | Active PAR; initiated 2023-02-14 | Defines energy-efficiency metrics for quantum computing (gate-based, annealing, simulation), including control chains and NISQ vs fault-tolerant devices. Relevant to ESG/“green IT” reporting when AIS environments rely on quantum backends. |
| P3185 | Standard for Hybrid Quantum-Classical Computing | Active | Defines hardware/software architecture for hybrid quantum-classical environments and APIs between QPUs and CPUs/GPUs/TPUs/FPGAs.Directly relevant to AIS and audit analytics architectures that integrate quantum accelerators as services. |
| P1943 | Standard for Post-Quantum Network Security | Active | Specifies methods to implement post-quantum versions of existing network security protocols, including hybrid modes and handling larger PQC keys. Provides a standards-based template for hardening AIS network |

| Standard | Title/Description | Status | Notes |
|---|---|---|---|
| | | | channels against CRQCs without defining new algorithms. |
| P3172 | Recommended Practice for Post-Quantum Cryptography Migration | Active; late-stage draft, widely referenced by 2025 | Recommended practice / roadmap for PQC migration: inventory cryptographic assets, assess quantum risk, prioritize systems, and implement crypto-agility. Highly relevant for AIS migration planning and aligns conceptually with NIST PQC FIPS and ETSI QSC migration reports. |
| P7130 | Standard for Quantum Technologies Definitions | Active (approved 2021) | Establishes common definitions for quantum technologies, improving terminological consistency across IEEE standards and with ISO/IEC JTC 3. Supports your institutional argument about vocabulary convergence. |
| P7131 | Standard for Quantum Computing Performance Metrics & Benchmarking | Active PAR | Defines performance metrics and benchmarking methods for quantum computing; complements P3329 by focusing on performance rather than energy.Relevant to how firms interpret vendor claims in disclosures and audit evidence. |
| P2995 | Trial-Use Standard for a Quantum Algorithm Design and Development | Active trial-use | Provides processes for quantum algorithm design and development lifecycle. Indirect AIS link: development governance and documentation for quantum |

| Standard | Title/Description | Status | Notes |
|---|---|---|---|
| | | | routines embedded in financial analytics. |
| P1913 | YANG Model for Software-Defined Quantum Communication | Active | Defines YANG data models for SDN management of quantum links. Supports manageability and observability of quantum-enhanced networks that may carry AIS-relevant traffic. |

**Table 9: Institute of Electrical and Electronics Engineers (IEEE)**

| Standard | Title/Description | Status | Notes |
|---|---|---|---|
| EITCI Quantum Standards Group (QSG) | Quantum Standards Group landing page | Ongoing | Independent expert group hosted by EITCI Institute focusing on quantum information and communication technologies; aims to produce reference specifications that can feed into formal SDOs (ETSI, ISO/IEC, ITU).Current themes: quantum random number generation (QRNG) and quantum encryption protocols (OQP). |
| ONE-QUBIT PAD (OQP) Workgroup | One-Qubit Pad scheme workgroup | Ongoing | Develops the One-Qubit Pad (OQP), an entanglement-based protocol for encrypting quantum information using a single-qubit key, framed as a highly efficient "one-time pad" |

| Standard | Title/Description | Status | Notes |
|---|---|---|---|
| | | | analogue in the quantum domain. Provides protocol definitions, implementation guidance, and reference tests. |
| EITCI-QSG-OQP PROTOCOL | “Reference Standard for the One-Qubit Pad – Protocol (Definitions, Key Theoretical Concepts and Use Cases for Qubits)” | Version 0.1 (RFC-style) | Draft reference standard describing OQP terminology, key theoretical constructs, and example use cases; not a formal ISO/ETSI standard but an RFC-like technical baseline that can be proposed into other SDOs. |
| EITCI-QSG-OQP IMPLEMENTATION / Reference Standards Acceptance Vote | OQP implementation reference & acceptance procedure | Ongoing | Documents an exemplar OQP implementation and an internal “Reference Standards Acceptance Vote” mechanism used by QSG members to accept implementations as reference-conformant. Illustrates a bottom-up, community-driven standardization practice. |
| QRNG Workgroup (EQRNG) | Quantum Random Numbers Generation workgroup | Ongoing | Focuses on entanglement and non-entanglement-based QRNG schemes, with the explicit objective of initiating an international WG for a |

| Standard | Title/Description | Status | Notes |
|---|---|---|---|
| | | | consolidated QRNG technical specification and follow-on formal standardization. |
| EITCI-QSG-EQRNG TESTING | EQRNG testing framework | Ongoing | Defines testing methodologies for QRNG devices, including statistical and, potentially, physics-aware tests, intended as a reference for future QRNG technical standards. |
| EITC Certification Programme | EITC programme overview | Ongoing | Describes EITCI's broader ICT certification schemes; QSG output can be reflected in EITC certificates (e.g., for QRNG/OQP implementers), making QSG work visible via skills/certification rather than regulation. |
| EITCI Programme Committees | Programme committees page | Ongoing | Lists EITCI governance and expert committees, including those overseeing QSG activities, highlighting that QSG is an expert consortium rather than an intergovernmental SDO. |

**Table 10: European IT Certification Institute (EITCI)**

| Report | Source |
|---|---|

| Deloitte | https://www2.deloitte.com/us/en/insights/industry/technology/technology-media-and-telecom-predictions/2022/future-of-quantum-computing.html |
|---|---|
| Zapata computing | https://www.zapatacomputing.com/enterprise-survey/ |
| Statista | https://www.statista.com/study/115330/quantum-computing/ |
| Prophecy | https://www.prophecymarketinsights.com/market_insight/Insight/request-sample/571 |

**Table 11: Quantum Standards Supplements**

The provided sources (Table 11) are from various organizations and platforms, and they offer insights into the future of quantum computing and related market trends: Deloitte is a global consulting firm known for its research and insights into various industries, including technology. The report likely provides predictions and analyses on the future of quantum computing in the technology, media, and telecom sectors. Zapata Computing is a company specializing in quantum software and algorithms. Their enterprise survey report likely focuses on understanding the adoption, challenges, and opportunities of quantum computing in businesses and industries. Statista is a renowned platform for statistical data and market research. Their study on quantum computing is likely to provide comprehensive market insights, trends, and statistics related to the quantum computing industry. Prophecy Market Insights is a market research and consulting firm. Their report might offer detailed market analysis, trends, and growth projections for the quantum computing market.

| Organization | Source |
|---|---|
| ISO | https://www.iso.org/ |
| NIST | https://www.nist.gov/ |
| ETSI | https://www.etsi.org/ |
| QIRG | https://datatracker.ietf.org/group/qirg/ |
| ITU | https://www.itu.int/ |

| IEC | https://iec.ch/ |
|---|---|
| Cenelec | https://www.cencenelec.eu/ |
| GAO | https://www.gao.gov/products/gao-22-104422 |
| IEE | https://quantum.ieee.org/standards |
| BSI | https://www.bsigroup.com/en-GB/industries-and-sectors/quantum-technology/ |
| EITCI | https://eitci.org/technology-certification/qsg |
| NIH | https://www.ncbi.nlm.nih.gov/pmc/articles/PMC8995124/ |
| FCC | https://www.fcc.gov/news-events/events/2020/12/quantum-internet-forum |

**Table 12: Quantum Organizations**

Table 12 provides a consolidated reference list of the primary quantum standards organizations and their web portals. For AIS practitioners, this table serves as a practical index for locating current standards, migration guidance, and certification requirements relevant to quantum-ready AIS environments. Table 13 catalogs quantum-related reports published by the Government Accountability Office (GAO). While not a standard setting organization, these reports document federal oversight assessments of quantum computing risks, post-quantum cryptography migration needs, and policy coordination gaps that directly inform the institutional context within which AIS security and assurance obligations will evolve.

| GAO | Insite |
|---|---|
| GAO-22-104422 | https://www.gao.gov/products/gao-22-104422<br>Quantum Computing and Communications:<br>Status and Prospects<br>Oct 19, 2021. Publicly Released: Oct 19, 2021. |
| GAO-23-106826 | Cybersecurity: Launching and Implementing the National Cybersecurity Strategy<br><br>Jun 29, 2023. |
| GAO-23-106559 | Science & Tech Spotlight: Securing Data for a Post-Quantum World<br>Published: Mar 8, 2023 . |
| GAO-23-106571 | Priority Open Recommendations: Office of Science and Technology Policy<br>May 10, 2023 . |
| GAO-20-527SP | Science & Tech Spotlight: Quantum Technologies Published: May 28, 2020 . |

**Table 13: Government Accountability Office (GAO)**

- **GAO-22-104422 - Quantum Computing and Communications: Status and Prospects (Oct 19, 2021)**
  - o Assesses potential benefits and risks of quantum computing and quantum communications, and identifies policy options for Congress and agencies (e.g., supporting collaboration, workforce, investment, and supply chains) to foster development while mitigating security and economic risks.
  - o Highlights that quantum computers could transform fields like cryptography, drug discovery, and materials science, but are still years away from broad utility and require substantial investment and coordination.
- **GAO-23-106826 - Cybersecurity: Launching and Implementing the National Cybersecurity Strategy (Jun 29, 2023)**
  - o Reviews challenges in operationalizing the 2023 National Cybersecurity Strategy, including governance, information sharing, and emerging-technology threats.
  - o Recommends clearer roles, metrics, and oversight mechanisms for federal agencies, which indirectly shape how quantum and post-quantum risks are prioritized in U.S. cyber policy.
- **GAO-23-106559 - Science & Tech Spotlight: Securing Data for a Post-Quantum World (Mar 8, 2023)**
  - o Explains the “harvest-now, decrypt-later” threat and stresses the need for migration to post-quantum cryptography once standards are available.
  - o Identifies key policy issues: inventorying cryptographic dependencies, coordinating migration across sectors, and balancing security with implementation costs.

- **GAO-23-106571 - Priority Open Recommendations: Office of Science and Technology Policy (May 10, 2023)**
  - Summarizes GAO's open recommendations to Office of Science and Technology Policy (OSTP), including strengthening coordination on emerging technologies and research security (which covers quantum among other areas).
  - Calls for improved cross-agency planning and monitoring for science and technology initiatives, relevant to how quantum R&D and standards efforts are governed.
- **GAO-20-527SP - Science & Tech Spotlight: Quantum Technologies (May 28, 2020)**
  - Provides a concise policy brief on quantum computing, sensing, and communications, emphasizing potential national-security and economic impacts and highlighting uncertainties and long development timelines.
  - Flags policy issues such as research funding, workforce, standards, and international collaboration/export controls, many of which recur in later GAO work